\documentclass[a4paper,12pt]{article}
\usepackage{graphicx}
\def\half{\frac{1}{2}}

\def\beq{\begin{equation}}
\def\eeq{\end{equation}}
\def\bea{\begin{eqnarray}}
\def\eea{\end{eqnarray}}
\begin{document}
\title{A new approximation method for time-dependent problems in quantum
mechanics}
\author{Paolo Amore\thanks{paolo@ucol.mx}, Alfredo Aranda\thanks{fefo@ucol.mx}\\
{\small\it Facultad de Ciencias, Universidad de Colima,}
{\small\it Bernal D\'{\i}az del Castillo 340,}\\
{\small\it Colima, Colima, M\'exico}\\
Francisco M. Fern\'andez\thanks{fernande@quimica.unlp.edu.ar}\\
{\small\it INIFTA (Conicet,UNLP), Diag. 113 y 64 S/N, Sucursal 4,}\\
{\small\it Casilla de Correo 16, 1900 La Plata, Argentina}\\
Hugh Jones\thanks{h.f.jones@imperial.ac.uk}\\
{\small\it Department of Physics, Imperial College,}
{\small\it London SW7 2AZ, England}}
\date{}
\maketitle \vspace{-0.7cm}
\begin{abstract}
\noindent We propose an approximate solution of the time-dependent
Schr\"odinger equation using the method of stationary states
combined with a variational matrix method for finding the energies
and eigenstates. We illustrate the effectiveness of the method by
applying it to the time development of the wave-function in the
quantum-mechanical version of the inflationary slow-roll
transition.
\end{abstract}

{\small Pacs numbers {45.10.Db,04.25.-g}}\\

\newpage
There have been various approaches to the calculation of the time
development of the Universe in the early stages of inflation, in
which the inflaton field $\varphi$ evolves from the initial
unstable vacuum state in which $\langle\varphi\rangle=0$ to the
final stable vacuum in which $\langle\varphi\rangle=\pm a$, say.

At the level of quantum mechanics the earliest approach~\cite{CPI}
used the Hartree-Fock method, which, although not very accurate,
does give a qualitative picture of the true time development.
Several years later Cheetham and Copeland~\cite{CC} went beyond
the Gaussian approximation by using an ansatz which included a
second-order Hermite polynomial. This represented an improvement
on the Hartree-Fock approximation, but still did not reproduce the
first maximum in $\langle x^2 \rangle$ of the exact wave-function.
A further attempt to solve this difficult non-perturbative problem
involved the use of the linear delta expansion~\cite{HFJ}. This
was successful to the extent that it stayed close to the exact
solution for longer than previous methods, but there seemed to be
a barrier to its implementation beyond the first maximum. The
results of these various approaches to the calculation of $\langle
x^2 \rangle$ are shown in Fig.~1, where they are compared with the
exact result of Lombardo et al.~\cite{Lombardo}, obtained by
Fourier transforming to frequency space and then back again.

\begin{figure}[h!]
\begin{center}
\includegraphics[width=4in]{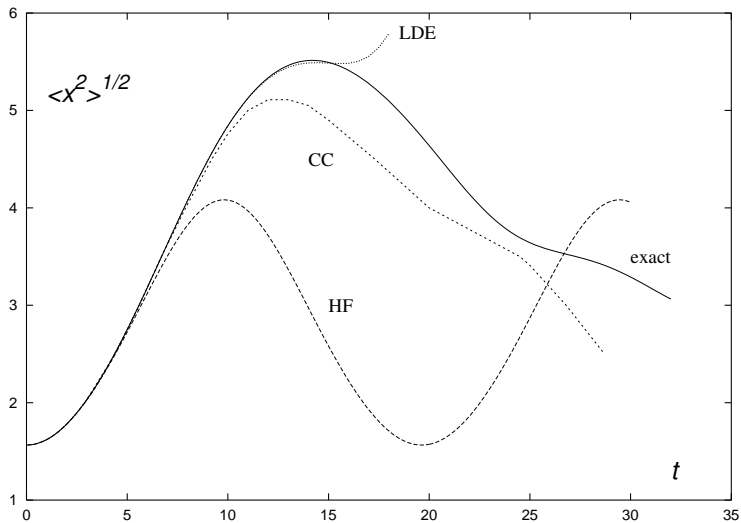}
\caption{\small $\langle x^2\rangle^{1/2}$ versus $t$ for the slow
roll potential of Eq.~(\ref{dwell}). The labels HF, CC and LDE
correspond to the Hartree-Fock method~\cite{CPI}, the improved
Hartree-Fock method ~\cite{CC} and the fourth-order linear delta
expansion~\cite{HFJ} respectively. These are compared with the
exact result of \cite{Lombardo}.} \label{Fig1}
\end{center}
\end{figure}

The present paper represents another approach to the problem,
using the method of stationary states, combined with a matrix
method for finding the energy eigenvalues of the full Hamiltonian,
together with their corresponding eigenfunctions. The method
consists of converting the eigenvalue equation into a matrix
equation by taking matrix elements with respect to harmonic
oscillator wave functions of arbitrary frequency $\Omega$ and
truncating the matrix at some finite size $N$. As
$N$ increases one may expect the method to become more and more
accurate for the lower-lying energy levels. However, for
arbitrary $\Omega$ the convergence is rather slow. A significant
component of our method is to choose an optimal
value of $\Omega$ by a particular
version of the principle of minimal sensitivity (PMS)\cite{PMS},
which greatly accelerates the convergence. The method has been
used before for the determination of the eigenvalues\cite{QM85},
but here we also make use of the corresponding wave-functions to
implement the method of stationary states to calculate the
time-dependence of the state from its initial configuration. It
turns out that this simple method is extremely accurate at even
quite small orders, and has no difficulty with the long-time
behaviour. We have concentrated on this particular problem to exemplify
the power of the method, but its range of application is obviously much
wider.

As stated above, we tackle the problem of solving the energy eigenvalue equation
\begin{eqnarray}
H \psi_n = E_n \psi_n \label{eq2}
\end{eqnarray}
by converting it to a matrix equation, using an orthonormal basis
of wave functions of the quantum harmonic oscillator, depending upon an arbitrary frequency
$\Omega\equiv \alpha^2$:
\begin{eqnarray}
\varphi_n(x) = N_n \ e^{- \alpha^2 x^2/2} \ H_n(\alpha x) \ ,
\label{basis}
\end{eqnarray}
with $N_n =(\alpha/(2^n \ n! \ \sqrt{\pi}))^\half$.
The slow-roll Hamiltonian involves a double-well potential, and has the specific form
\begin{eqnarray}
H&=&\half p^2 +
\lambda(x^2-a^2)^2/24 + const.\cr
&&\cr
&=& \half p^2 -\half m^2 x^2 + gx^4,
\label{dwell}
\end{eqnarray}
with  $m^2=\lambda a^2/6$ and $g=\lambda/24$.

Of course the infinite-dimensional matrix $H_{n\ell}$ must be truncated to some
finite dimension, say $N\times N$, and then its
eigenvalues can be calculated by simple matrix diagonalization.
A brute-force approach to the problem is to stick with a given $\Omega$
and rely on larger and larger values of $N$ to approach the desired
accuracy for the lower-lying energy levels. However, much improved
accuracy for even modest sized matrices can be obtained by a judicious
choice of $\Omega$. The criterion we shall adopt here, which
is essentially that adopted in \cite{QM85}, is the principle of
minimal sensitivity\cite{PMS} applied to the trace of the truncated
matrix.

The rationale behind this principle is that the eigenvalues and
other exact quantities of the problem are independent of $\Omega$
but any approximate result coming from the diagonalization method
for finite $N$ exhibits a spurious dependence on the oscillator
frequency. This also applies to the trace of the matrix, i.e. the
sum of the eigenvalues. A well-motivated criterion for choosing
$\Omega$ is therefore to take it at a stationary point of
$T_N\equiv \sum_{n=0}^{N-1} H_{nn}$, so that this independence is
respected locally. Thus we impose the PMS condition
\begin{eqnarray}
\frac{\partial}{\partial \Omega} {T}_N = 0 \ . \label{pms}
\end{eqnarray}
The reason for applying this condition to the trace is that
${T}_N$ is a simple quantity to evaluate, and moreover it is
invariant under the unitary transformation associated with a
change of basis. Once $\Omega$ is so determined, one obtains an
approximation to the first $N$ eigenvalues and eigenvectors of
$H$ by a numerical diagonalization of the truncated $N
\times N$ matrix. One could also contemplate applying the PMS to
the determinant, which of course shares the property of invariance
under unitary transformations, but this would be a much more
cumbersome calculation, and could well introduce many spurious PMS
solutions.

In order to implement the method we need the harmonic oscillator
matrix elements of $x^p$. Closed formulas have been given in
Ref.~\cite{Morales}, which we adapt here for completeness.
\beq
(x^{2r})_{n\ell}=\frac{\sqrt{n!\ell!}}{(2\alpha)^{2r}}\sum_{k=0}^{{\rm min}(n,r-\lambda)}\frac{(2r)!}
{2^{2r-k-\lambda}(r-\lambda-k)!(n-k)!(2\lambda+k)!k!}
\eeq
for $\ell-n=2\lambda$, and
\bea
\hspace{-2cm}(x^{2r+1})_{n\ell}&=&\frac{\sqrt{n!\ell!}}{(2\alpha)^{2r+1}}
\sum_{k=0}^{{\rm min}(n,r-\lambda)}\cr && \hspace{1.5cm}\frac{(2r+1)!}
{2^{2r-k-\lambda+\half}(r-\lambda-k)!(n-k)!(2\lambda+1+k)!k!}
\eea
for $\ell-n=2\lambda+1$. These formulas assume $\ell\ge n$, but the matrix
is symmetric. In both cases $r$ must be greater than $\lambda$. For all other
values of $p$, $n$ and $\ell$ the matrix elements vanish.

In addition to these we will need the matrix elements of $p^2$, which are
given by
\beq
(p^2)_{n \ell} = -\frac{1}{2}\alpha^2\left[\sqrt{\ell(\ell-1)}\delta_{\ell,n+2}-(2\ell+1)\delta_{\ell,n}
+\sqrt{(\ell+1)(\ell+2)}\delta_{\ell,n-2}\right].
\eeq
Using these results we can write down the matrix elements of
Eq.~(\ref{dwell}) in the basis of the wave functions of
Eq.~(\ref{basis}). The trace to order $N$ is given by \beq
\frac{4}{N}{T}_N =
N(\Omega+\frac{m^2}{\Omega})+\frac{g}{\Omega^2}( 1+2N^2), \eeq
which turns out to have a single minimum, located at \beq
\Omega_{PMS} = \frac{1}{3}X^{1/3}-\frac{m^2}{X^{1/3}}, \eeq where
\beq X \equiv \frac{3}{N} \left[ 9g( 1+2N^2) + {\sqrt{3}}
\left(N^2m^6+27g^2(1+2N^2)^2\right)^\frac{1}{2}\right] \eeq

\begin{figure}[h!]
\begin{center}
\includegraphics[width=4in]{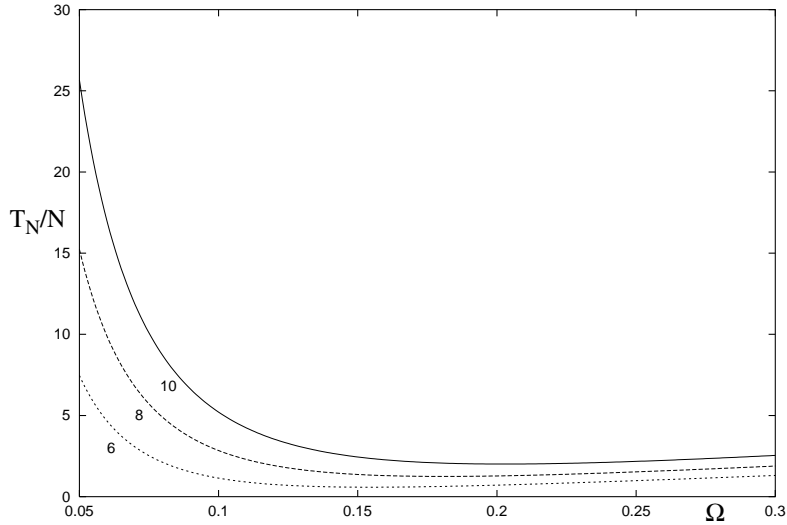}
\caption{\small Trace of the truncated matrix for the
Hamiltonian~(\ref{dwell}) normalized by the number of states as a
function of the variational parameter $\Omega$ for $a=5$,
$\lambda=0.01$. The three curves are for $N=6,\,8,\,10$. }
\label{fig1n}
\end{center}
\end{figure}

A graph of ${T}_N/N$ against $\Omega$ is given in Fig.~\ref{fig1n}
for various values of $N$. Here we have taken the parameters as
$a=5$, $\lambda=0.01$, in accordance with previous work on the
subject\cite{CPI, CC, HFJ}. By taking $\Omega$ at the minimum, in
accordance with PMS, we obtain a good approximation to the
spectrum for even quite small matrices.

For the present purposes, the important point of this method is that in addition to the
energy levels it gives an approximation to the energy eigenfunctions, in the form
\begin{eqnarray}
\psi_n(x) = \sum_{k=0}^{N-1} {d}_{nk}  \ \varphi_k(x) \ .
\label{eq5}
\end{eqnarray}
where $d_{nk}$ denotes the $k$th component of the $n$th eigenvector of the  truncated
Hamiltonian matrix.

These can be used to implement the method of stationary states.  Namely, if the
initial wave-function at $t=0$ is expanded as
\begin{eqnarray}
\Psi(x, 0) = \sum_{n=0}^\infty a_n \psi_n(x) \label{Psi0},
\end{eqnarray}
its value at a later time is given  by
\begin{eqnarray}
\Psi(x,t) = \sum_{n=0}^\infty a_n \ e^{-i E_n t} \ \psi_n(x)
\label{Psit)}
\end{eqnarray}
As an intermediate step we expand $\Psi(x, 0)$ in terms of the eigenvectors $\varphi_n(x)$:
\begin{eqnarray}
\Psi_n(x, 0) = \sum_{k=0}^{N-1} c_n  \ \varphi_n(x) \ .
\label{cn}
\end{eqnarray}
By comparison with Eq.~(\ref{Psi0}) we see that
\begin{eqnarray}
a_n = \sum_{\ell=0}^{N-1} c_\ell \ ({\bf d}^{-1})_{\ell n} \ .
\label{an}
\end{eqnarray}
Notice that by this stage $\Omega$ has been determined by the PMS condition, so that the
matrix inversion here is numerical, rather than symbolic.

It remains to determine the coefficients $c_n$ in Eq.~(\ref{cn}).
The initial wave function used in previous studies of slow-roll
inflation is given by
$$
\Psi(x,0) = \left(\frac{m}{\pi} \right)^{1/4}
e^{- m x^2/2}.
$$
By orthonormality,
\begin{eqnarray}
c_n = \int \phi_n^*(x) \Psi(x,0) dx = N_n \ \left(\frac{m}{\pi} \right)^{1/4} \
\int_{-\infty}^{+\infty}  e^{- \beta^2 x^2/2} H_n(\alpha x) dx,
\end{eqnarray}
where $\beta^2 \equiv m + \alpha^2$. By means of the
change of variable $y = \alpha x$
\begin{eqnarray}
c_n = \frac{N_n}{\alpha} \left(\frac{m}{\pi} \right)^{1/4} \int_{-\infty}^{+\infty}
e^{- \frac{\beta^2}{2 \alpha^2} y^2} H_n(y) dy \ .
\end{eqnarray}
and the expansion of $H_n(y)$, namely
\begin{eqnarray}
H_n(y) &=& \sum_{k=0}^{[n/2]} (-1)^k \frac{n!}{(n-2 k)! k!} 2^{n-2 k} y^{n -2 k} \ .
\end{eqnarray}
we obtain
\begin{eqnarray}
c_n &=& \frac{N_n}{\alpha} \left(\frac{m}{2\pi} \right)^{1/4}
\sum_{k=0}^{[n/2]} (-1)^k \frac{n!}{(n-2 k)! k!} 2^{n-2 k} \ J_k,
\end{eqnarray}
where
\begin{eqnarray}
J_k &\equiv& \int_{-\infty}^{+\infty}   y^{n -2 k}  e^{- \frac{\beta^2}{2 \alpha^2} y^2} dy
= \left( \frac{\sqrt{2} \alpha}{\beta} \right)^{n-2 k+1} \ \int_{-\infty}^{+\infty} z^{n-2 k} \ e^{-z^2} dz \ . \nonumber
\end{eqnarray}
In fact $c_n = 0$ unless $n$ is even $n = 2 \ell$, and then
\begin{eqnarray}
c_{2 \ell} &=& \frac{N_{2 \ell}}{\alpha}  \left(\frac{m}{\pi} \right)^{1/4}
\sum_{k=0}^{\ell} (-1)^k \frac{(2 \ell)!}{(2 \ell-2 k)! k!} 2^{2 (\ell -k)}
\left( \frac{\sqrt{2} \alpha}{\beta} \right)^{2 (\ell - k)+1}
\Gamma\left(\ell-k+ \frac{1}{2} \right).\nonumber\cr\\&&
\end{eqnarray}
The coefficients $a_n$ are now given by Eq.~(\ref{an}).

\begin{figure}[h!]
\begin{center}
\includegraphics[width=4in]{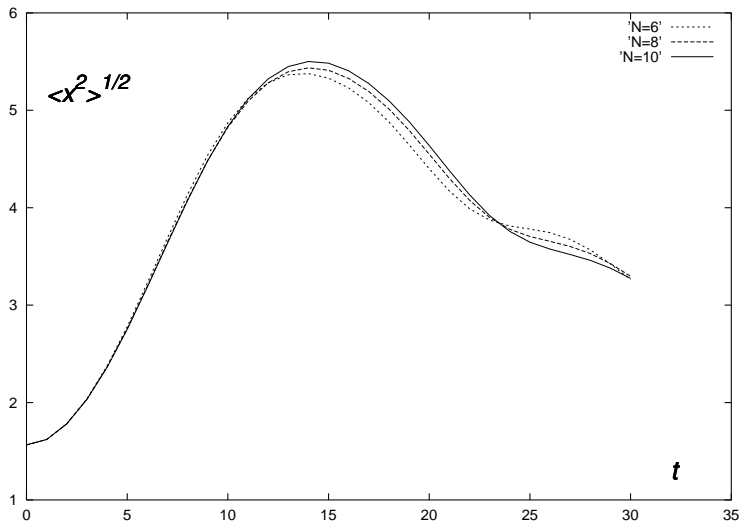}
\caption{$\langle x^2\rangle^{1/2}$ versus $t$ for the slow roll
potential of Eq.~(\ref{dwell}) calculated with our method for $N$=6, 8 and 10.
On the scale of the figure, the result for $N$=10 is barely distinguishable from
the  exact result.}
\label{Fig3}
\end{center}
\end{figure}

For comparison with previous work, we use our time-dependent wave
function to calculate $\langle x^2\rangle$, given by
\begin{eqnarray}
\langle x^2 \rangle &=&  \int \Psi^*(x,t)  \ x^2 \ \Psi(x,t) \ dx
\end{eqnarray}
assuming that $\Psi(x,t)$ remains normalized, which we have
checked. In terms of the $\psi_n(x)$ this is
\begin{eqnarray}
\langle x^2 \rangle &=&  \sum_{n , \ell}  \ a_n^* \ a_\ell \ e^{-
i \omega_{n\ell} t} \ \int \psi_n(x) \ x^2 \ \psi_\ell(x) \ dx,
\end{eqnarray}
where $\omega_{n\ell} \equiv E_n - E_\ell$. Using Eq.~(\ref{eq5})
this becomes \beq \langle x^2\rangle =  \sum_{n , \ell}  \ a_n^* \
a_\ell \sum_{k,j} d_{nk} d_{\ell j} (x^2)_{kj} \ \cos
\omega_{n\ell} t \ . \eeq The result for $\langle x^2
\rangle^{1/2}$ is plotted in Fig.~\ref{Fig3}. As can be seen, the
method is vastly superior to the previous approximative methods,
and by $N=10$ the curve can hardly be distinguished from that
obtained using Fourier transform methods.
Although for the present purposes a basis of size 10 was
sufficient, considerably greater accuracy for both energies and
eigenfunctions can readily be obtained by going to larger $N$.
This costs little additional effort because, once the PMS has been
applied, the Hamiltonian matrix is fully numerical and the
calculation of its eigenvalues/eigenvectors can be calculated
numerically with accuracy and speed. For example, with $N=100$ we
obtain the energy of the ground state of the quartic oscillator
Hamiltonian $p^2+x^2+2gx^4$ with $g=1000$ to an accuracy of 58
significant figures. Although we have no analytic proof, our
numerical work strongly suggests that the error decreases
exponentially with $N$.

In both these cases we have been dealing with symmetric
potentials, but with a slight modification it can easily be
extended to asymmetric potentials. In such a case, instead of
using a basis of harmonic oscillator wave-functions centered on the origin,
we can take them to be centered on a shifted position
$x=\sigma$. The method then contains two variational parameters
$\Omega$ and $\sigma$, which are to be determined by the PMS
condition that $T_N$ be stationary in both.

Another possible extension of the method might be to use the
eigenfunctions of the $q$-deformed harmonic oscillator as a basis instead
of those of the simple harmonic oscillator. A variational method
using these eigenfunctions has been shown to give very
accurate results for the ground-state energy of the anharmonic
oscillator\cite{Kim}.

In conclusion, we have shown that the matrix method, combined with
the principle of minimal sensitivity applied to the trace of the
Hamiltonian matrix, is a powerful tool for finding the spectrum of
arbitrary polynomial potentials having only bound states. Since
the method also gives the energy eigenfunctions it is ideal for
implementing the method of stationary states in order to track the
time development of a given initial wave function. For quite
modest-sized matrices the method gives good accuracy, even for
long time scales, as we have demonstrated for the slow-roll
potential.

\vspace{12pt}

\noindent P.A. acknowledges support of Conacyt grant no. C01-40633/A-1.

\noindent A.A. acknowledges support from Conacyt grant no. 44950 and PROMEP.

\end{document}